\documentclass[twocolumn,secnumarabic,amssymb, nobibnotes, aps]{revtex4-1}

\usepackage{graphicx}
\usepackage{dcolumn}
\usepackage{bm}

\usepackage{color}
\usepackage{braket}

\newcommand{\rr}{{\bf r}}

\begin{document}

\begin{abstract}
A reconfigurable logic gate is proposed in a two-dimensional double quantum wire system with a coupling window enabled by a Rashba field. Manipulating the spin states of incoming electrons several quantum logic gates (OR, AND, XOR, CNOT) can be implemented. The logic gate functionality can be switched by tuning the Rashba parameter only. In this context, we investigate suitable configurations of the device region by embedding a quantum point contact located in the coupling region to obtain all four logic gates. The ballistic spin polarized transmission functions are calculated using an effective mass scattering formalism in the framework of a multi-channel, multi-terminal system. Owing its versatility, the proposed logic gate can be integrated in programmable architectures, able to implement both classical and quantum algorithms.   
\end{abstract}

\title{Reconfigurable quantum logic gates using Rashba controlled spin polarized currents}

\author{G. A. Nemnes$^{1,2,*}$ and Daniela Dragoman$^{1,3}$}
\affiliation{$^1$University of Bucharest, Faculty of Physics, MDEO Research Center, 077125 Magurele-Ilfov, Romania}%
\affiliation{$^2$Horia Hulubei National Institute for Physics and Nuclear Engineering, 077126 Magurele-Ilfov, Romania}%
\affiliation{$^3$Academy of Romanian Scientists, Splaiul Independentei 54, Bucharest 050094, Romania}%

\email{E-mails: nemnes@solid.fizica.unibuc.ro}

\maketitle

\section{Introduction}

Although it is currently in the early stages of development, the field of quantum computing is rapidly gaining ground \cite{Ladd2010,ref10.1103/RevModPhys.90.015002}. Certain problems, like integer factorization and the simulation of quantum many-body problems, with applications to cryptography and device modeling, can be resolved in a shorter amount of time by quantum computers. However, presently, quantum computers are thought to have a great impact only on a relatively restricted class of problems as compared to the classical counterparts.
Therefore it is expected that, in general, quantum computing would complement classical computing and both types of approaches may coexist in future architectures. 
 For example, a quantum algorithm like Shor's algorithm for factoring numbers can be approached in two steps: in a first step, the problem is formulated as finding the period of a function, a task which can be achieved by a classical computer, and, in a second step, a significant speedup is brought by quantum Fourier transforms.

Decoherence effects, which are mainly due to thermal fluctuations and interactions with other dephasing elements (e.g. impurities with degrees of freedom, etc), are the major obstacle in operating large number of qubits. Approaches which tackle this problem successfully include trapped ions \cite{PhysRevA.84.030303,Schafer2018}, semiconductor quantum dots \cite{Veldhorst2014,Veldhorst2015,1674-1056-27-2-020305}, superconducting circuits \cite{Barends2014} or single photons \cite{RevModPhys.79.135}. Several implementations of classical or quantum logic gates (LGs) as single devices have been proposed so far using ballistic transport, from both theoretical and experimental point of view. Such examples include serial quantum point contacts for multi-valued logic gates \cite{Seo2014}, double quantum waveguide system with a coupling window \cite{0953-8984-27-1-015301}, quantum logic gates based on ballistic transport in graphene \cite{doi:10.1063/1.4943000}, nanowire quantum point contact registers \cite{doi:10.1021/acs.nanolett.6b01840}, ballistic deflection transistors \cite{Wolpert2011,1742-6596-647-1-012066}, which may be also implemented as graphene Y-junctions \cite{0957-4484-29-35-355202,0953-8984-30-44-445301}. Other concepts for computation are being considered such as valleytronic-based information processing \cite{PhysRevB.96.245410}.

A recurring geometry is that of two coupled quantum wires. The coupling region can be tuned by magnetic fields, in which case the operation of C-NOT \cite{doi:10.1063/1.1405428} and $\sqrt{\rm NOT}$ \cite{0953-8984-27-1-015301} logic gates can be reproduced. The C-NOT gate operation can be achieved also using two quantum wires coupled through a potential barrier \cite{PhysRevLett.84.5912}, while a NOT-gate was implemented in coupled GaAs-AlGaAs quantum wires \cite{1278281}.   

The spin degree of freedom is regarded as a feasible option for a robust implementation of qubits. Manipulation of the electron spin in nanoelectronic devices is typically achieved by electric field control exerted through Rashba spin-orbit interaction. Several device geometries have been considered for spin-filtering such as cross-junctions \cite{PhysRevLett.83.376}, T-junctions \cite{doi:10.1063/1.1347023}, Y-junctions \cite{doi:10.1063/1.2364859} or, more recently, triangular networks of quantum nanorings \cite{DEHGHAN201821}.

We propose here a reconfigurable logic gate able to implement several operation modes, switching between AND, OR, XOR and CNOT gates. The device structure is essentially a double quantum waveguide system, where the coupling is achieved by a Rashba field, which tunes the functionality of the logic gate. Previously, the two parallel waveguides were coupled capacitively \cite{PhysRevB.82.085311}, by tunneling \cite{PhysRevB.76.195301} or by an external magnetic fields \cite{doi:10.1063/1.1405428,0953-8984-27-1-015301}. We investigate device structures which enable the switching between the five aforementioned gates by tuning only the Rashba parameter. The versatility of the proposed logic gate can result in novel computing architectures, able to switch from classical to quantum algorithms.

\section{Model and methods}

\subsection{The device model}

\begin{figure}[t]
\centering
\includegraphics[width=8cm]{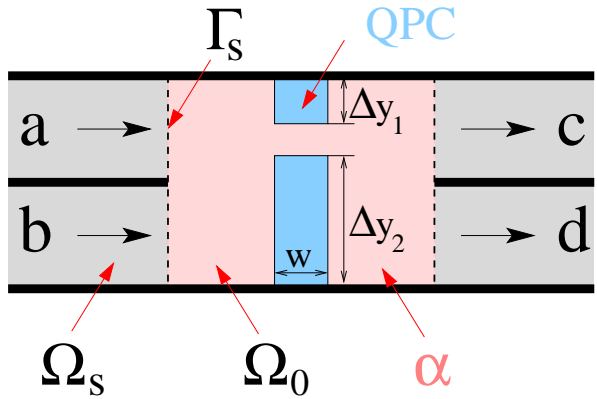}
\caption{Device geometry: two incoming (a,b) and two outgoing (c,d) leads corresponding to regions $\Omega_s$ are connected to a coupling window $\Omega_0$ of size $L_x \times L_y$. The interfaces are denoted by $\Gamma_s = \Omega_s \cap \Omega_0$. The Rashba spin-orbit interaction is present in $\Omega_0$, with a constant coupling parameter $\alpha$. A quantum point contact (QPC) is superimposed, assembled by two barriers of sizes $\Delta y_1 \times w$ and $\Delta y_2 \times w$ and height $V_0$, symmetrically positioned along the transport direction.}
\label{dev}
\end{figure}

The device structure is a two-dimensional double quantum wire system with a rectangular coupling window, where the top gate electric field generates a uniform Rashba field, as depicted in Fig. \ \ref{dev}. It contains two input leads, labeled $a$ and $b$, and two output leads, $c$ and $d$. The typical effect of the spin-orbit interaction region is that of a quantum well, where, in addition, a mixing between the spin dependent sub-bands takes place. In this context, an upward shift in energy introduced e.g. by potential barriers would partly compensate the Rashba field.

The geometry of the system is specified by the dimensions of the coupling region $L_x = 4$ $\mu$m, $L_y = 1$ $\mu$m, and device material parameters such as the effective mass of InAs $m^* = 0.023$ $m_0$ and Rashba coupling parameter $\alpha$, which can be tuned by the top gate electric field up 20 meV nm. Here we investigate a device structure with a point contact (QPC) embedded in the coupling region in two configurations: symmetric and asymmetric with respect to the $a$ and $b$ inputs, this aspect being essential in the realization of the logic gates. The two barrier regions assembling the QPC have the lateral sizes $\Delta y_1$ and $\Delta y_2$, the thickness along the transport direction $w$ and the potential energy $V_0$. The four leads are identical with a square-well potential in the transversal direction.

\subsection{The scattering formalism}

The ballistic transport in the double quantum wire system is described in the framework of a multi-channel, multi-terminal scattering formalism. The transmission functions are calculated efficiently for a large energy set using the R-matrix method, which was initially developed by Wigner and Eisenbud in the field of nuclear physics \cite{PhysRev.72.29}. As the scattering theory became relevant for the field of nanoelectronics, the R-matrix method was employed in the description of the coherent transport properties in mezoscopic systems \cite{SMRCKA1990221,PhysRevB.63.085319,PhysRevB.58.16209}. The R-matrix method was applied to ballistic nanotransistors \cite{doi:10.1063/1.1748858,doi:10.1063/1.2113413,doi:10.1063/1.3269704}, thermoelectric effects in quantum wires \cite{NEMNES20101613}, spin transport in Datta-Das type transistors \cite{1742-6596-338-1-012012}, and, more recently in the description of time-dependent transport and snaking states in core-shell nanowires \cite{doi:10.1021/acs.nanolett.6b01840,PhysRevB.93.205445}.

The system is partitioned in two incoming leads, the scattering region which coincides with the coupling window and two outgoing leads. The scattering problem is formulated as the time-independent Schr\"odinger equation 
\begin{equation}
\label{sch}
{\mathcal H}\Psi(\rr) = E\Psi(\rr),
\end{equation}  
with asymptotic boundary conditions for the four leads, which are discussed in the next sub-section. 
The Hamiltonian of the system in the scattering region $\Omega_0$ contains the scattering potential $V(\rr \in \Omega_0)$ and the Rashba term: 
\begin{equation}
\label{H0}
{\mathcal H_0}=-\frac{\hbar^2}{2m^*}\bigtriangleup + V(\rr \in \Omega_0) + \frac{\alpha}{\hbar}(p_y\sigma_x-p_x\sigma_y), 
\end{equation}
where $\alpha$ is Rashba coupling parameter. 
The Hamiltonian inside the leads $\Omega_s$ includes the confinement potential $V_s(\rr \in \Omega_s)$:
\begin{equation}
\label{Hs}
{\mathcal H_s}=-\frac{\hbar^2}{2m^*}\bigtriangleup + V_s(\rr \in \Omega_s). 
\end{equation}

The wavefunctions inside the leads can be written as general solutions of the Hamiltonians ${\mathcal H}_s$:
\begin{eqnarray}
\label{Psis}
\Psi_s(\rr \in \Omega_s;E)
&=& \sum_i  \Psi_{\nu}^{\rm in} \exp{(-ik_{\nu} z_s)}
     \ket{\Phi_\nu(\rr^\perp_s)} \nonumber\\
    &+& \sum_i \Psi_{\nu}^{\rm out} \exp{(ik_{\nu} z_s)}
        \ket{\Phi_\nu(\rr^\perp_s)}, 
\end{eqnarray}
where $k_{\nu} = \sqrt{ {2m^* \over \hbar^2}(E-E_{\perp}^{\nu}) }$ are the wavevectors
along the transport direction in each channel $\nu$.
The composite index $\nu \equiv (s,i,\sigma)$ denotes the channel $i$ with spin $\sigma$ from lead $s$. The transversal modes in each lead are in general superpositions of spin states:
\begin{equation}
\label{Phi}
\ket{\Phi_\nu(\rr^\perp_s)} = \sum_\sigma \Phi_{\nu}(\rr^\perp_s) \ket{\sigma},
\end{equation}
where $\ket{\sigma}$ are the eigenfunctions of the spin operator $S_z$.

The R-matrix is defined as:
\begin{equation}
{\rm R}_{\nu\nu'}(E) = -\frac{\hbar^2}{2} \sum_{l=0}^{\infty}  
                             \frac{(\chi_{l})_\nu(\chi_{l}^*)_{\nu'}}{E-E_{l}},
\end{equation}
with
\begin{equation}
\label{chilnu}
(\chi_{l})_\nu = \int_{\Gamma_s} d \Gamma_s \Phi_\nu(\rr^\perp_s) \chi_{l,\sigma}(\rr\in\Gamma_s).
\end{equation}

The functions and energies $\ket{\chi_{l}} = \sum_\sigma \chi_{l,\sigma} \ket{\sigma}$ and $E_{l}$ are obtained by solving the Wigner-Eisenbud problem, which is independent on the total energy $E$:
\begin{eqnarray}
\label{we1}
& &{\mathcal H_0} \ket{\chi_l} = E_l \ket{\chi_l},\\
\label{we2}
& &\;\;\;\left[ \frac{\partial \chi_{l,\sigma}}{\partial z_s} \right]_{\Gamma_s} = 0.
\end{eqnarray}

The scattering S-matrix is calculated based on the R-matrix \cite{doi:10.1063/1.1748858} at a given energy $E$:
\begin{equation}
\label{smatrix}
{\rm S} = - \left[ {\rm 1} - \frac{i}{m^*} {\rm R} {\rm k} \right]^{-1} \left[ {\rm 1} + \frac{i}{m^*} {\rm R} {\rm k} \right]
\end{equation}
where ${\rm k}$ is a diagonal matrix, ${\rm k}_{\nu\nu'}=k_\nu\delta_{\nu\nu'}$. It relates the output coefficients to the input coefficients in Eq.\ (\ref{Psis}): $\vec{\Psi}^{\rm out} = {\rm S} \vec{\Psi}^{\rm in}$. 

If the electrons are incoming from one channel $\nu'$, the normalized input coefficients are $\Psi_{\nu''}^{in} = \delta_{\nu''\nu'}$ for any $\nu''$ and the transmission function into channel $\nu$ at given energy $E$ can be determined from the unitary symmetric matrix 
$\tilde{\rm S}={\rm k}^{1/2}{\rm S}{\rm k}^{-1/2}$ as 
${\mathcal T}_{\nu}(E) = |\tilde{\rm S}_{\nu\nu'}(E)|^2$. 
The total spin-dependent transmission in terminal $s$, as the electrons are incoming from all other terminals $s' \neq s$ is obtained by summing up incoherent contributions from all incoming transversal modes $\nu'$ into the transversal modes $i$ in terminal $s$: $T_{s,\sigma} = \sum_{i,\nu'} |\tilde{\rm S}_{\nu\nu'}(E)|^2$.
In general, a coherent superposition of incoming channels, in this case from terminals $a$ and $b$, may be considered.

\subsection{Asymptotic conditions and transmission symmetries}

In the 4-terminal system with $a$, $b$ input terminals and $c$, $d$ output terminals, the input logic states $S = A, B, C, D$ are defined for each transversal mode as:
\begin{eqnarray}
\label{input1}
A &\equiv& \frac{1}{\sqrt{2}} \ket{\uparrow}_a + \frac{1}{\sqrt{2}} \ket{\uparrow}_b, \\
B &\equiv& \frac{1}{\sqrt{2}} \ket{\uparrow}_a + \frac{1}{\sqrt{2}} \ket{\downarrow}_b, \\
C &\equiv& \frac{1}{\sqrt{2}} \ket{\downarrow}_a + \frac{1}{\sqrt{2}} \ket{\uparrow}_b, \\
\label{input4}
D &\equiv& \frac{1}{\sqrt{2}} \ket{\downarrow}_a + \frac{1}{\sqrt{2}} \ket{\downarrow}_b,
\end{eqnarray}
which correspond to (a,b) inputs: (0,0), (0,1), (1,0), (1,1). All four terminals have identical sets of transversal eigenmodes.

Assuming a single transversal mode in each lead, we consider a coherent superposition of two incoming channels, one from lead $a$ and one from $b$, corresponding to the lowest eigenmode ($i=0$) and the spin-dependent transmission in the output lead $s = c, d$ may be written as:
\begin{equation}
T_{s,\sigma} =  \left| \frac{1}{\sqrt{2}} \tilde{\rm S}_{\nu\nu'} + 
                                  \frac{1}{\sqrt{2}} \tilde{\rm S}_{\nu\nu''} \right|^2,
\end{equation}
where $\nu \equiv (s, 0, \sigma)$, $\nu' \equiv (a, 0, \sigma')$ and $\nu'' \equiv (b, 0, \sigma'')$. Here, $\sigma'$ and $\sigma''$ are set by the input states defined in Eqs.\ (\ref{input1}-\ref{input4}). 
 
For a symmetric QPC in the scattering region, the transmission function is invariant under the exchange of the output leads and/or spin orientation, depending on the input $S$ and the orientation of the Rashba field. 

For $B$ and $C$ input states, by changing $\alpha \rightarrow -\alpha$, the transmission functions obey the following symmetry operations:
\begin{eqnarray}
\label{BC-minus}
& &T^B_{c\uparrow}(\alpha) = T^B_{d\downarrow}(-\alpha), \;\;\;
T^B_{c\downarrow}(\alpha) = T^B_{d\uparrow}(-\alpha), \nonumber\\
& &T^C_{c\uparrow}(\alpha) = T^C_{d\downarrow}(-\alpha), \;\;\;
T^C_{c\downarrow}(\alpha) = T^C_{d\uparrow}(-\alpha).
\end{eqnarray}
For $A$ and $D$ input states, we have:
\begin{equation}
\label{AD-minus}
T^A_{c\uparrow}(\alpha) = T^D_{d\downarrow}(-\alpha), \;\;\;
T^A_{c\downarrow}(\alpha) = T^D_{d\uparrow}(-\alpha)
\end{equation}

In addition, for the same $\alpha$ values, 
\begin{equation}
\label{same-alpha-BC}
T^B_{c\downarrow}(\alpha) = T^C_{d\downarrow}(\alpha), \;\;\;
T^B_{d\uparrow}(\alpha) = T^C_{c\uparrow}(\alpha).
\end{equation}
and
\begin{eqnarray}
\label{same-alpha-AD}
& &T^A_{c\uparrow}(\alpha) = T^D_{d\downarrow}(\alpha), \;\;\;
T^A_{d\uparrow}(\alpha) = T^D_{c\downarrow}(\alpha), \nonumber\\
& &T^A_{c\downarrow}(\alpha) = T^D_{d\uparrow}(\alpha), \;\;\;
T^A_{d\downarrow}(\alpha) = T^D_{c\uparrow}(\alpha).
\end{eqnarray}

A graphical representation of these symmetries is given in Figs.\ S1 and S2 in the Supplementary Material (SM).

\section{Results and discussion}

The electrons prepared in the quantum states $S = A, B, C, D$ incoming coherently from the input leads $a$ and $b$ are subject to quantum interference in the scattering region, leading to spin polarized currents in the output leads $c$ and $d$. 
We define the spin polarization $p_s$ in the output terminals $s = c, d$, considering linear bias regime as:
\begin{equation}
p^S_s = \frac{T^S_{s\uparrow} - T^S_{s\downarrow}}{T^S_{s\uparrow} + T^S_{s\downarrow}}. 
\end{equation}
Depending on the sign of the output spin polarization $p_s$ a logical $0$ or $1$ is assigned. We impose a threshold value of 10\% for a net spin current to deliver a proper signal. In order to distinguish between the realization of different logic gates, in the following we shall denote by LGI the logic gate index: AND/NAND = $\pm$1, OR/NOR = $\pm$2, XOR/XNOR = $\pm$3 and CNOT = 4. 

The aim is to develop LGs with tunable functionality, by changing only one parameter, which can be externally controlled, in this case the Rashba parameter $\alpha$. This depends on the electric field perpendicular to the transport direction, induced by applying a top gate voltage. Accordingly, we may vary $\alpha$ in the interval $[-\alpha_{\rm max},\alpha_{\rm max}]$ to render different LGs. In this context, we consider the geometrical device structure parameters are fixed, along with the Fermi energy $E_{\rm F}$ and QPC potential $V_0$.

We consider both symmetrical and asymmetrical embedded QPCs, the latter enabling the realization of CNOT. Our approach is to test several geometrical configurations and for each assess, for a given energy $E$, the realization of a particular LG by varying $\alpha$. Then, by overlapping the results from all these instances we identify the energy values suitable to achieve two or more LGs and attribute one of these values to the Fermi energy of the system. Furthermore, at the selected energy, the $\alpha$ values corresponding to a certain LGI are readily available.

In the following we provide a proof-of-concept and the methodology to identify multi-functional LGs configurations, leaving the detailed optimization beyond the scope of the present paper.

\subsection{Symmetric QPC}

We first analyze the case of a symmetric quantum point contact, labeled as SYM-QPC. In this case, the spin polarizations in the two output terminals, $c$ and $d$, obey the relations:
\begin{eqnarray}
\label{ps-sym-prop}
& &p^B_c(\alpha) = -p^B_d(-\alpha), \;\;\;
p^C_c(\alpha) = -p^C_d(-\alpha), \nonumber\\
& &p^A_c(\alpha) = -p^D_d(-\alpha), \;\;\;
p^D_c(\alpha) = -p^A_d(-\alpha). 
\end{eqnarray}
We note that for SYM-QPC, only the functionality of the classical gates (AND, OR, XOR and the inverted outputs) can be reproduced as these have (a,b) to (b,a) symmetry, unlike CNOT.

\begin{figure}[t]
\centering
\includegraphics[width=8cm]{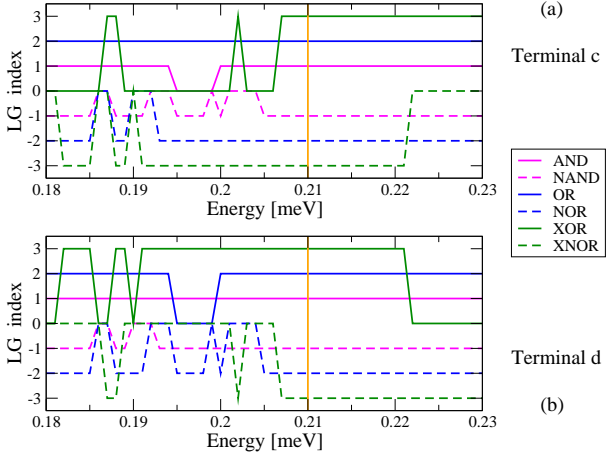}
\caption{Logic gate overview for SYM-QPC: output for $c$ and $d$ terminals on a selected energy range. Note that in the case of SYM-QPC, XOR$_c$ $\equiv$ XNOR$_d$ and XOR$_d$ $\equiv$ XNOR$_c$, while AND/NAND and OR/NOR gates do no present this symmetry. At the energy $E=0.21$ meV all 6 logic gates are obtained for distinct $\alpha$ values.}
\label{lgo-sym}
\end{figure}

The SYM-QPC is defined by $\Delta y_1$ = $\Delta y_2 = L_y/4$, $w = L_x/8$ and $V_0=1$ eV, which corresponds to almost perfectly opaque barriers. An overview of a typical LG realization is shown in Fig.\ \ref{lgo-sym}, where the LGI is represented versus the energy of incoming electrons, on a selected energy range. Transmission functions are calculated for $N_E= 250$ energies, equally spaced in the interval $[0,E_{\rm max}]$, with $E_{\rm max} = 0.25$ meV, and $N_\alpha = 200$ values for the spin-orbit coupling parameter, in the interval $[-\alpha_{\rm max},\alpha_{\rm max}]$, with $\alpha_{\rm max} = 20$ meV nm. This requires the solutions of the $N_E \times N_\alpha$ two-dimensional scattering problems, which are efficiently handled in parallel. For each instance, the four input values, corresponding to in-states $A, B, C, D$ in terminals $a$ and $b$ are connected with the output results, which are the spin polarizations $p_s$ in terminals $c$ and $d$. If the response function corresponds to a certain aforementioned LG, a finite LGI is assigned, otherwise LGI is set to zero.
Fig.\ \ref{lgo-sym} shows that for $E=0.21$ meV it is possible to obtain all six classical gates, for different values of $\alpha$. Moreover, this is achieved for both output terminals $c$ and $d$. One should note that for the XOR gate output in terminal $c$ corresponds to the XNOR gate output in terminal $d$ and the opposite is also true. However, it is not the case for AND and OR gates.

\begin{figure}[t]
\centering
\includegraphics[width=8cm]{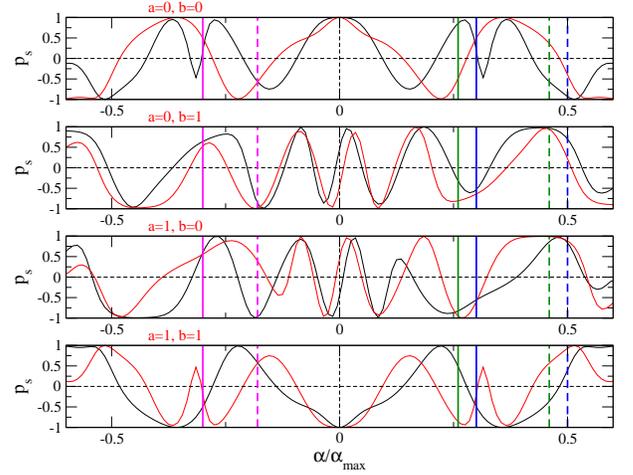}
\caption{Spin polarization $p^S_s$ in terminals $c$ (black) and $d$ (red). The vertical lines correspond to logic gates AND/NAND (magenta), OR/NOR (blue) and XOR/XNOR (green) obtained for terminal $c$, at $E=0.21$ meV, as indicated in Fig.\ \ref{lgo-sym}. Solid lines correspond to AND, OR, XOR logic gates, while the dashed lines indicate their negated counterparts.}
\label{ps-sym}
\end{figure}

\begin{figure}[t]
\centering
\includegraphics[width=8cm]{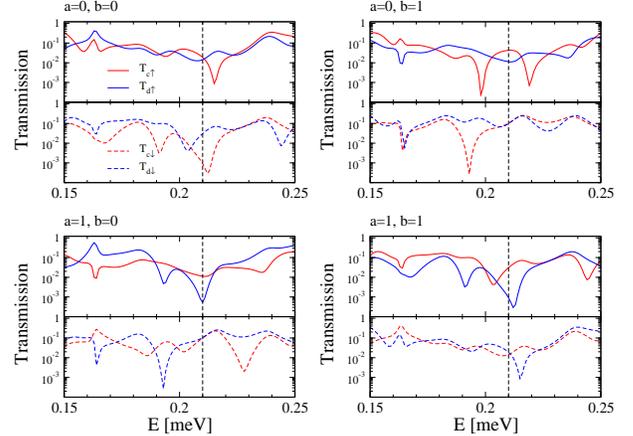}
\caption{Transmission functions for SYM-QPC for $\alpha = 5.3$ meV nm, for all four input values. At the energy $E=0.21$ meV, marked by vertical dashed lines, the XOR gate is obtained. The symmetry properties indicated by Eqs.\ (\ref{same-alpha-BC}) and (\ref{same-alpha-AD}) are readily verified.}
\label{trans-sym}
\end{figure}

To gain more insight regarding the realization of the different classical gates, we plotted in Fig.\ \ref{ps-sym} the spin polarization as functions of $\alpha$ in terminals $c$ and $d$, corresponding to the four possible inputs $(a,b)$, for $E=0.21$ meV. The $\alpha$ values corresponding to one of the AND, OR and XOR LGs, and their negated counterparts, are marked by vertical lines, for visibility only for the terminal $c$. Evidently, the indicated $\alpha$ values are not unique, but one $\alpha$ value corresponds to a single LG realization. A few observations and comments are at hand. The symmetry properties of $p_s$ shown in Eqs.\ (\ref{ps-sym-prop}) are consistently reproduced. For the input states $A$ and $D$, which correspond to up-up and down-down spins, and for $\alpha=0$, $p_s$ takes the values +1 and -1. This is not the case for $B$ and $C$ states, where the scattering process leads to smaller values for $p_s$. In addition for $A$ and $D$ states we have additional symmetry conditions, $p^A_s(\alpha) = p^A_s(-\alpha)$ and $p^D_s(\alpha) = p^D_s(-\alpha)$, for both terminals $c$ and $d$.

Increasing $\alpha$ an oscillatory behavior is found for $p_s$ in both output terminals. The spin precession is enhanced, which leads to an alternation in the collected spin states, cycling from up-spin to down-spin. The values of $p_s$ reach the maximal values ($\pm$ 1) and are de-phased for terminals $c$ and $d$. This indicates that indeed very high spin polarizations may be achieved for single LG realization. However, in general, invoking the constrain of multiple LG realization, the actual $p_s$ values are lower.

Figure\ \ref{trans-sym} shows a typical set of spin-dependent transmission functions corresponding to both outlet terminals $c$ and $d$, obtained for $\alpha = 5.3$ meV nm, which are calculated for each of the four inputs. At the energy of 0.21 meV this corresponds to XOR gate, as indicated in Fig.\ \ref{ps-sym}. The transmissions functions corresponding to AND and OR gates are represented in Figs.\ S3 and S4 in the SM, respectively. The signatures of resonant transport through the Rashba coupling region with embedded QPC are visible, as the transmission functions vary 2-3 orders in magnitude. The symmetries of the transmission functions reflected in in Eqs.\ (\ref{same-alpha-BC}) and (\ref{same-alpha-AD}) are clearly visible. Switching between $B$ and $C$ input states, the transmission for a spin component $\sigma$ remains the same as the terminals $c$ and $d$ are swapped. Interchanging $A$ and $D$ input states and swapping the terminals $c$ and $d$, the spin-dependent transmission functions are equivalent only if the spin is also changed.

\subsection{Asymmetric QPC}

\begin{figure}[t]
\centering
\includegraphics[width=8cm]{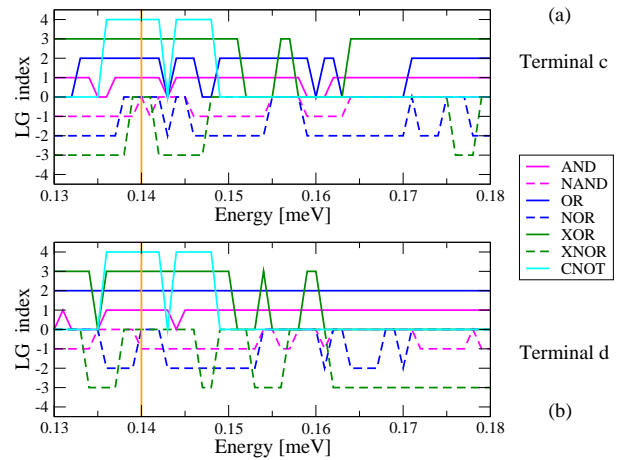}
\caption{Logic gate overview for ASYM-QPC. For a selected energy $E=0.14$ meV all 4 logic gates are obtained in both output terminals ($c$ and $d$) for different $\alpha$ values: AND, OR, XOR and CNOT.}
\label{lgo-asym}
\end{figure}

\begin{figure}[t]
\centering
\includegraphics[width=8cm]{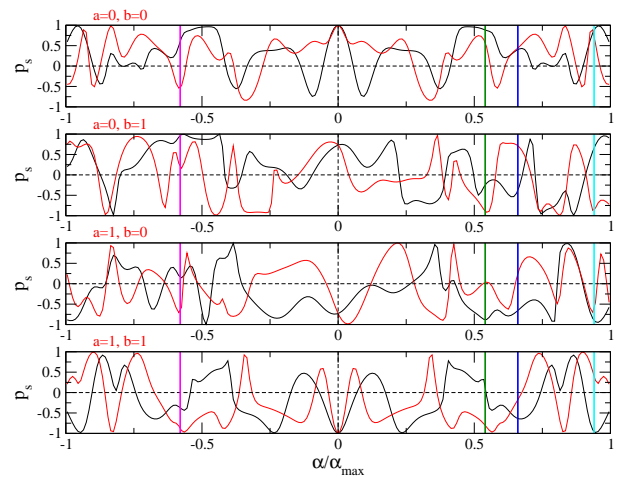}
\caption{Spin polarization $p^S_s$ in terminals $c$ (black) and $d$ (red), for $E=0.14$ meV. The vertical lines correspond to logic gates AND (magenta), OR (blue), XOR (green) and CNOT (cyan). The classical gates are indicated for terminal $c$, while for CNOT both $c$ and $d$ output values are observed.}
\label{ps-asym}
\end{figure}

Next we investigate the possibility to add the functionality of quantum LGs, CNOT in particular.
A distinctive aspect in comparison to the classical gates, is that a correlated output form both $c$ and $d$ terminals is required. 
As noted before, a symmetric QPC cannot reproduce CNOT behavior in the chosen representation of the input states. For example state $B$ should be related to the output (0,1), while state $C$ should give (1,1). Obviously, in a symmetric system this cannot be achieved. Several possibilities for introducing an asymmetry may be considered, which may be here divided in two classes, which correspond to a non-uniform map of the electrostatic potential or of the Rashba field.

We pursue the first option and consider an asymmetric quantum point contact (ASYM-QPC) specified by $\Delta y_1 = 0$, $\Delta y_2 = 0.75 L_y$, $w = L_x/8$ and $V_0 = 1$ eV. Applying the same procedure as in the case of SYM-QPC, we monitor the realization of the AND, OR, NOR and CNOT LGs at given energy. Figure\ \ref{lgo-asym} depicts an energy interval where all four gates appear. 

Selecting $E = 0.14$ meV, the spin polarizations in the output terminals $p_s$ are indicated in Fig.\ \ref{ps-asym}. There are several similitudes and also differences with respect to SYM-QPC. For $A$ and $D$ states the $\alpha \rightarrow -\alpha$ symmetry is retained, and the spin polarizations are maximized for $\alpha = 0$. However, in this case, Eqs.\ (\ref{ps-sym-prop}) are not anymore valid. 
The asymmetric potential produces larger variations between up-spin and down-spin in the two output terminals, as it is visible for low values of $\alpha$.

\begin{figure}[t]
\centering
\includegraphics[width=8cm]{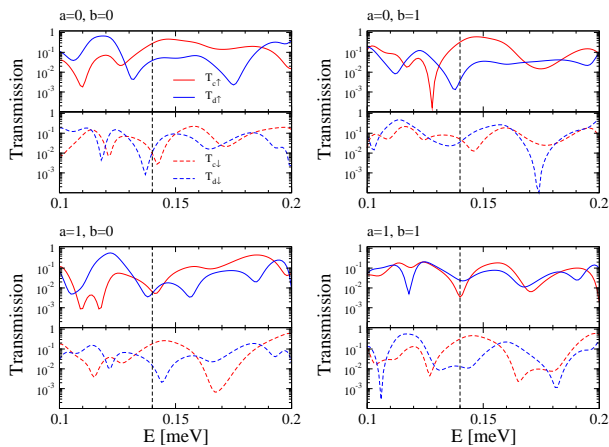}
\caption{Transmission functions for ASYM-QPC for $\alpha = 18.9$ meV nm. At the energy $E=0.14$ meV, marked by vertical dashed lines, the CNOT gate is obtained.}
\label{trans-asym}
\end{figure}

The spin-dependent transmission functions corresponding to CNOT realization are indicated in Fig.\ \ref{trans-asym}. This occurs for $\alpha = 18.9$ meV nm and $E = 0.14$ meV. In contrast to Fig.\ \ref{trans-sym}, the symmetry properties, reflected by Eqs.\ (\ref{same-alpha-BC}) and (\ref{same-alpha-AD}) are not anymore reproduced. Regarding the $\alpha \rightarrow -\alpha$ symmetry, only Eqs.\ (\ref{AD-minus}) are fulfilled.
Thus, we demonstrated the appearance of all four gates at a given energy, which may be chosen as the Fermi energy in the device structure.

\section{Conclusions}

We investigated the prospects for achieving a reconfigurable logic gate behavior in a double quantum wire with a Rashba coupling region. In the 4-terminal system the Rashba field together with an embedded quantum point contact mix the spin channels, rendering both classic and quantum logic gate behavior. The main focus here was to show that the logic gate functionality can be switched by a single external parameter, in this case, the Rashba parameter, without changing the structural configuration. We demonstrated that several classic gates (AND, OR, XOR) and their counterparts with negated outputs can be achieved at given Fermi energy and pre-set device geometry, using symmetric profiles of the electrostatic potential and Rashba interaction. By contrast, a quantum CNOT logic gate behavior can be reproduced using an asymmetric device structure, which adds up to the classical gates. Our results suggest that optimized configurations of reconfigurable logic gates can be achieved and employed in programmable architectures, which can implement both classical and quantum algorithms. \\

{\bf Acknowledgments} \\

This work was supported by a grant of Ministery of Research and Innovation, CNCS-UEFISCDI, project number PN-III-P4-ID-PCE-2016-0122, within PNCDI III.



\bibliography{manuscript}

\onecolumngrid

\newpage 

\appendix*
\section{Supplementary Material}

\renewcommand{\thefigure}{S\arabic{figure}}
\setcounter{figure}{0}

\begin{figure*}[h]
\centering
$T^B_{c\uparrow}(\alpha) = T^B_{d\downarrow}(-\alpha)$ \hspace*{3cm}
\includegraphics[width=8cm]{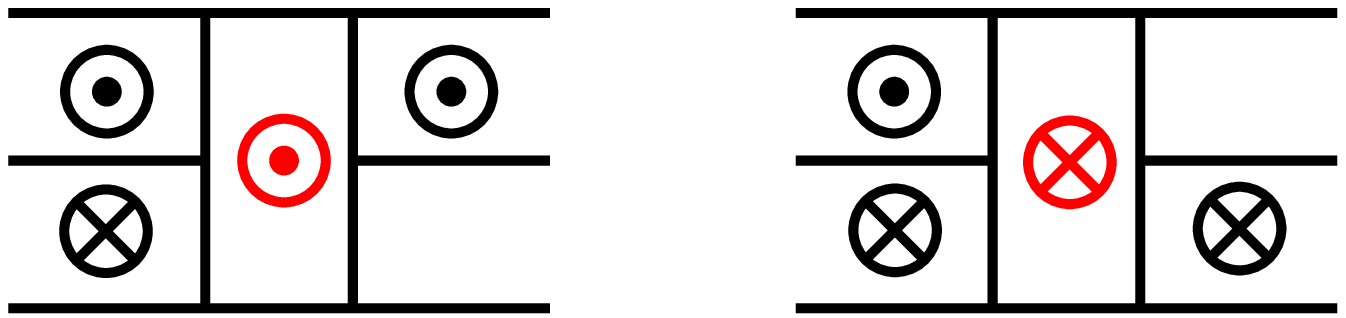} \vspace*{1cm}\\
$T^B_{c\downarrow}(\alpha) = T^B_{d\uparrow}(-\alpha)$ \hspace*{3cm}
\includegraphics[width=8cm]{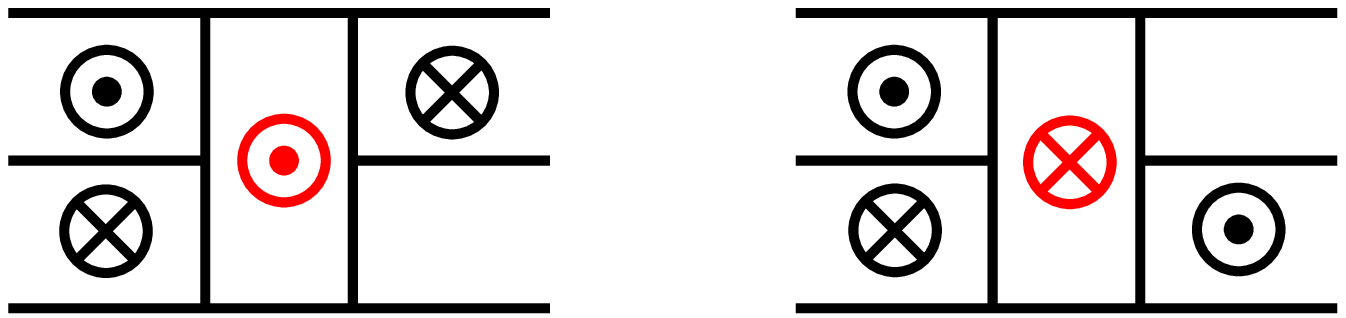} \vspace*{1cm}\\
$T^C_{c\uparrow}(\alpha) = T^C_{d\downarrow}(-\alpha)$ \hspace*{3cm}
\includegraphics[width=8cm]{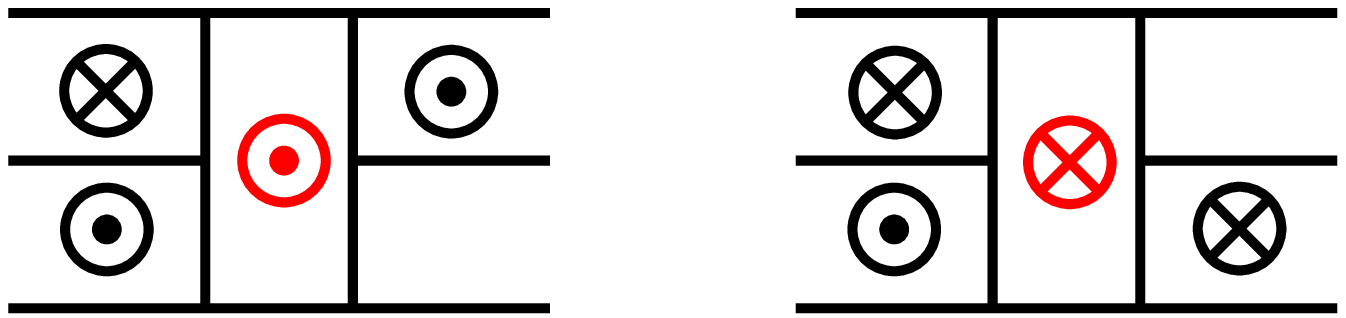} \vspace*{1cm}\\
$T^C_{c\downarrow}(\alpha) = T^C_{d\uparrow}(-\alpha)$ \hspace*{3cm}
\includegraphics[width=8cm]{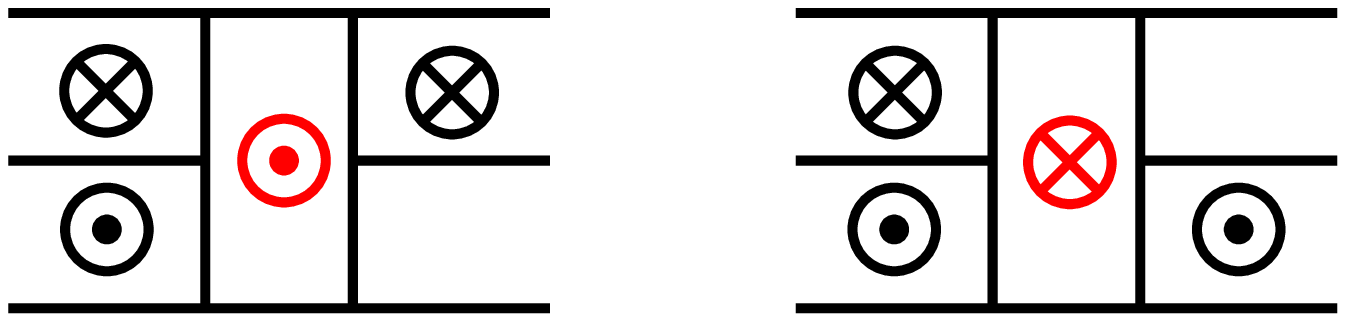} \vspace*{1cm}\\
$T^A_{c\uparrow}(\alpha) = T^D_{d\downarrow}(-\alpha)$ \hspace*{3cm}
\includegraphics[width=8cm]{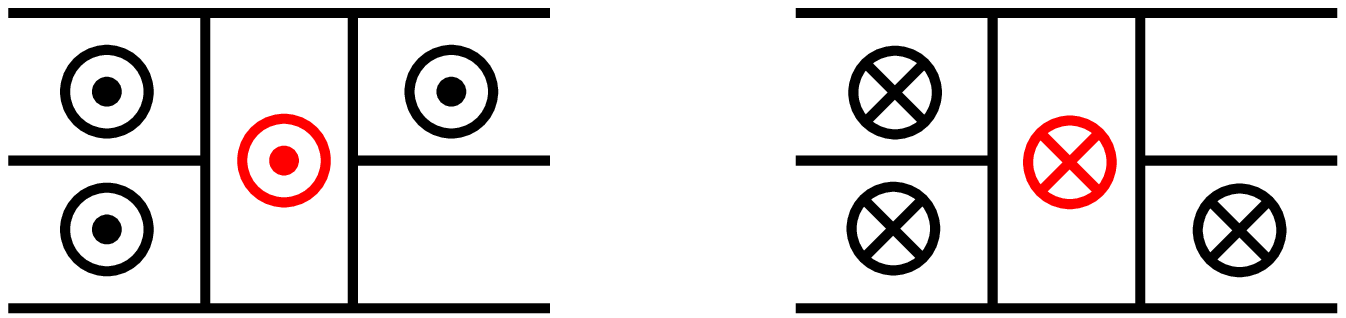} \vspace*{1cm}\\
$T^A_{c\downarrow}(\alpha) = T^D_{d\uparrow}(-\alpha)$ \hspace*{3cm}
\includegraphics[width=8cm]{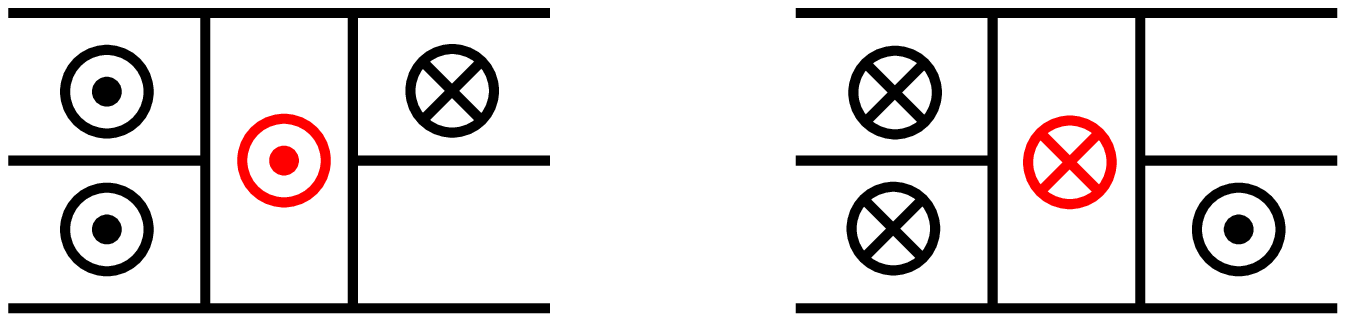} \vspace*{1cm}\\
\caption{Symmetry properties of transmission functions under inversion of the electric field inducing Rashba coupling ($\alpha \rightarrow -\alpha$). The electron spins and the electric field are denoted by black and red symbols, respectively.}
\label{a}
\end{figure*}

\begin{figure}[t]
\centering
$T^B_{c\downarrow}(\alpha) = T^C_{d\downarrow}(\alpha)$ \hspace*{3cm}
\includegraphics[width=8cm]{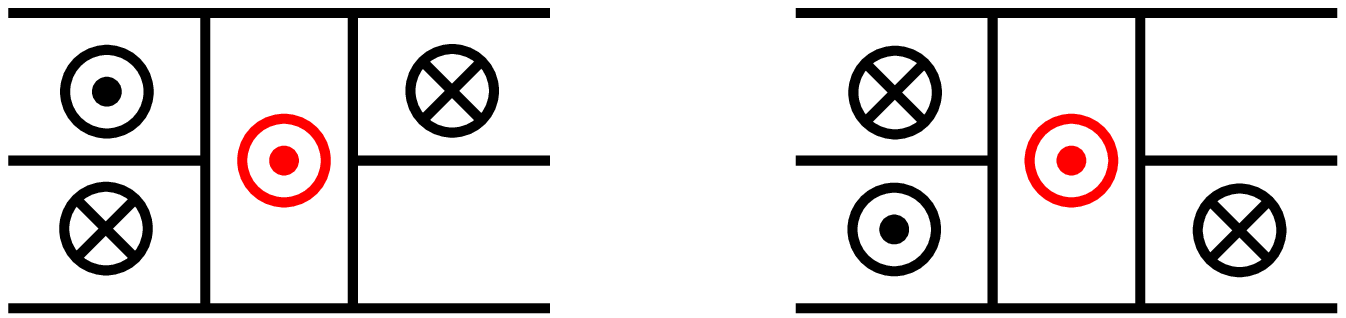} \vspace*{1cm}\\
$T^B_{d\uparrow}(\alpha) = T^C_{c\uparrow}(\alpha)$ \hspace*{3cm}
\includegraphics[width=8cm]{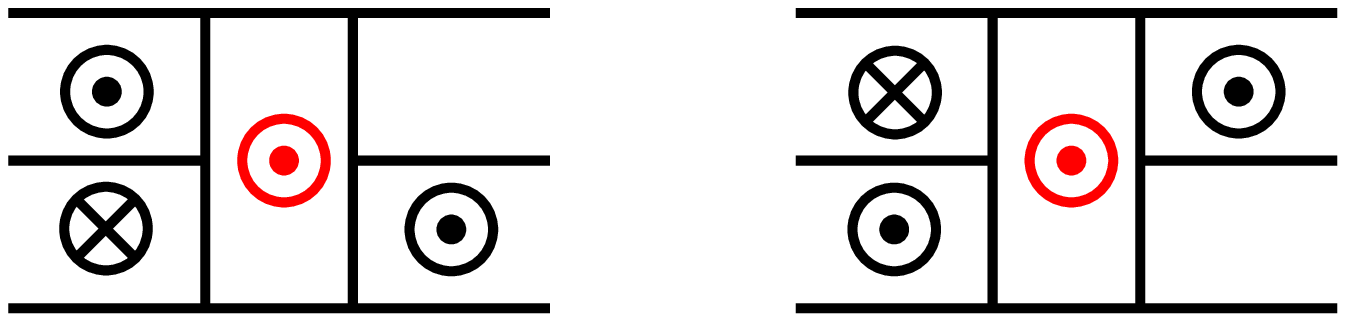} \vspace*{1cm}\\
$T^A_{c\uparrow}(\alpha) = T^D_{d\downarrow}(\alpha)$ \hspace*{3cm}
\includegraphics[width=8cm]{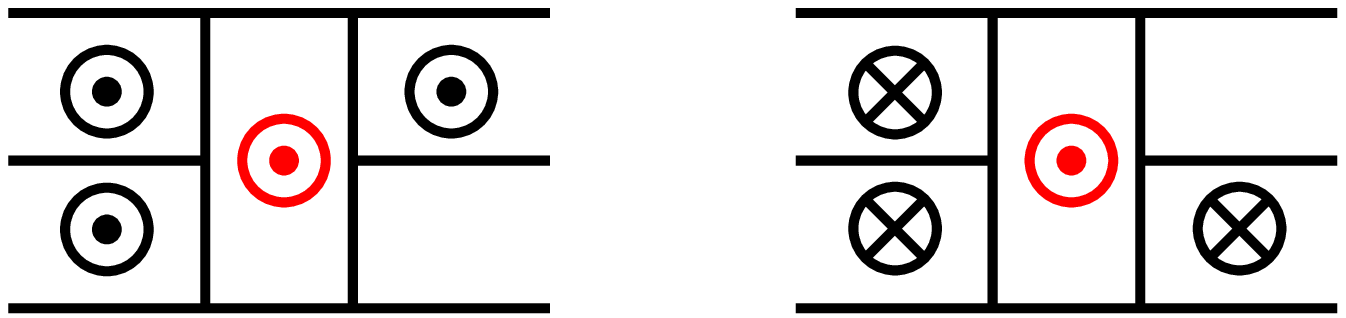} \vspace*{1cm}\\
$T^A_{d\uparrow}(\alpha) = T^D_{c\downarrow}(\alpha)$ \hspace*{3cm}
\includegraphics[width=8cm]{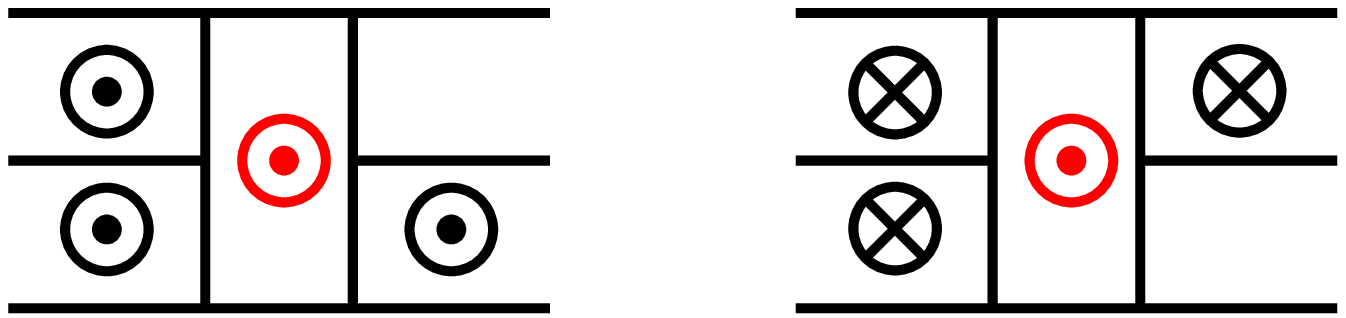} \vspace*{1cm}\\
$T^A_{c\downarrow}(\alpha) = T^D_{d\uparrow}(\alpha)$ \hspace*{3cm}
\includegraphics[width=8cm]{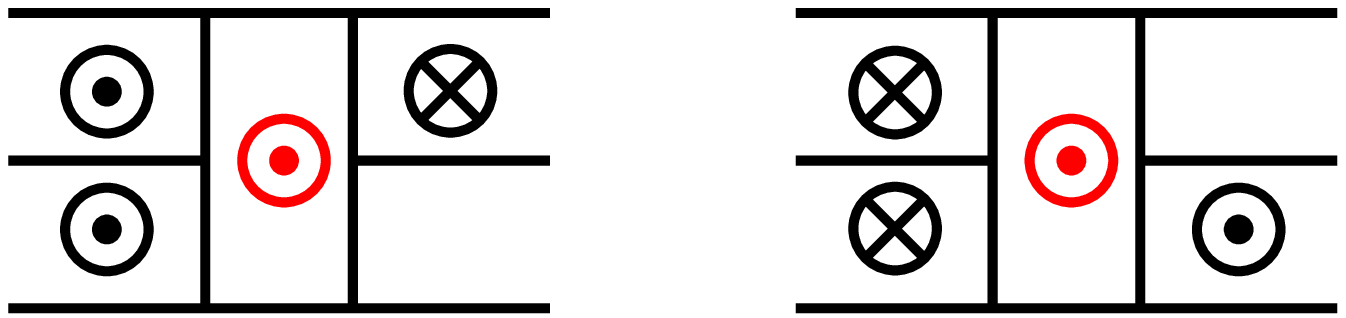} \vspace*{1cm}\\
$T^A_{d\downarrow}(\alpha) = T^D_{c\uparrow}(\alpha)$ \hspace*{3cm}
\includegraphics[width=8cm]{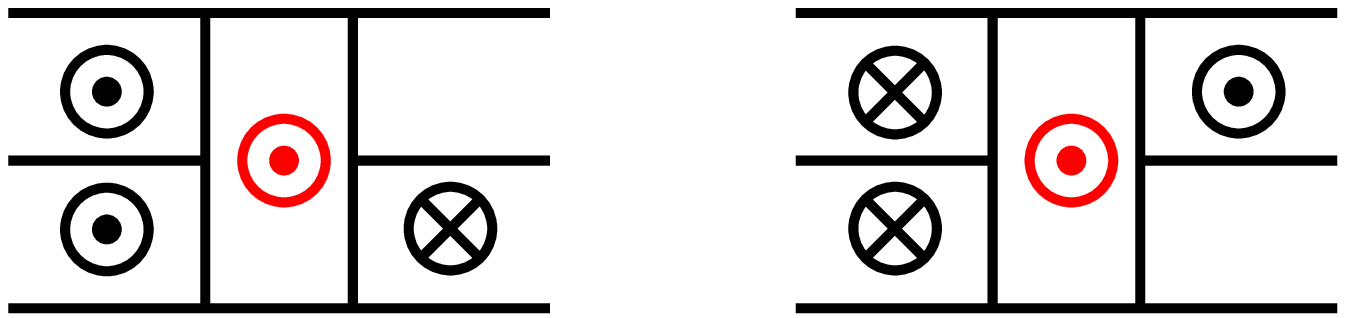} \vspace*{1cm}\\
\caption{Symmetry properties of transmission functions for a symmetric scattering potential in the coupling region for a given Rashba parameter $\alpha$. The electron spins and the electric field are denoted by black and red symbols, respectively.}
\label{b}
\end{figure}

\begin{figure}[t]
\centering
\includegraphics[width=14cm]{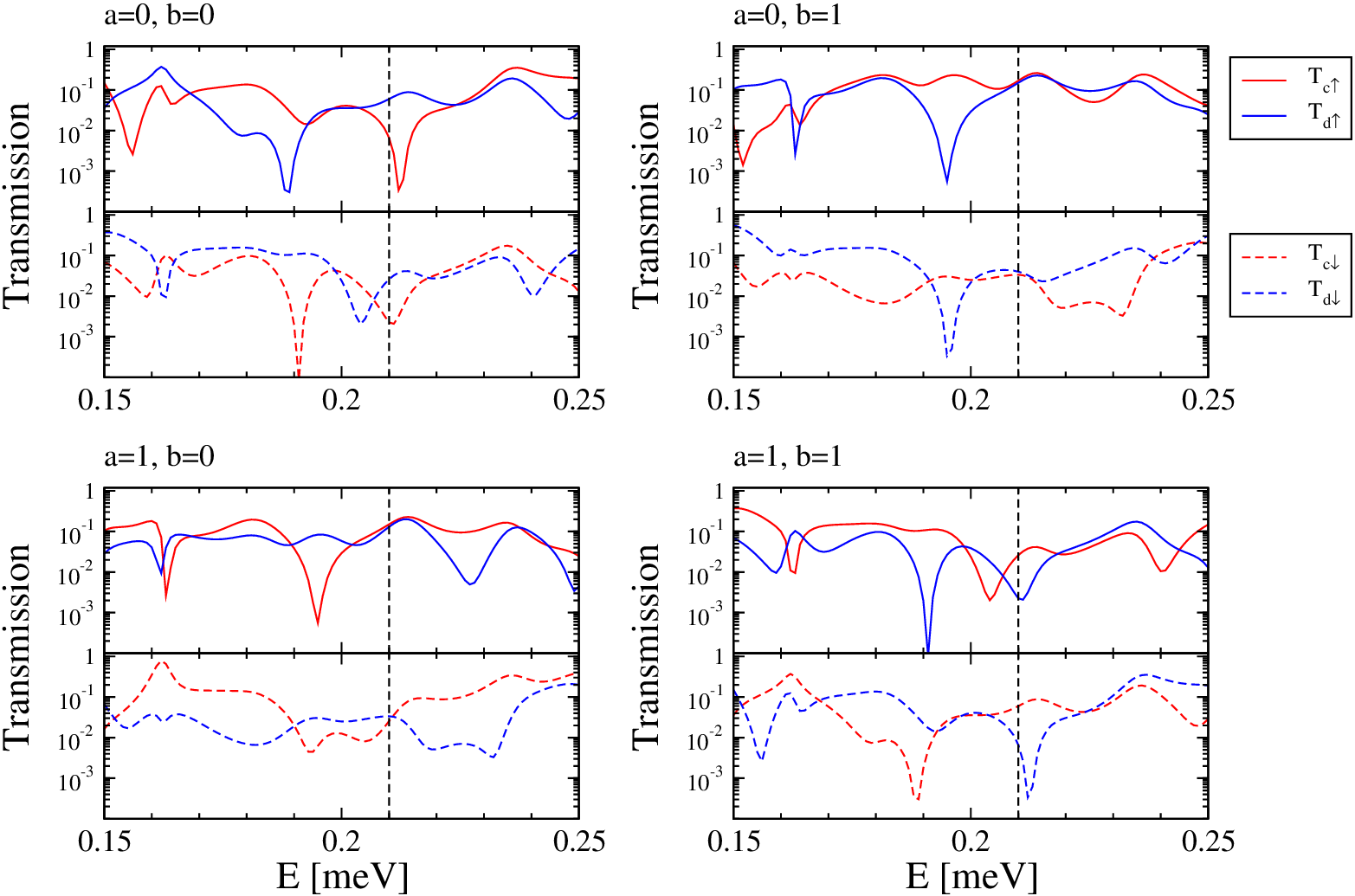} 
\caption{Transmission functions for a symmetric QPC showing the realization of an AND gate, at $E=0.21$ meV and $\alpha = -5.9$ meV nm.}
\label{and}
\end{figure}

\begin{figure}[t]
\centering
\includegraphics[width=14cm]{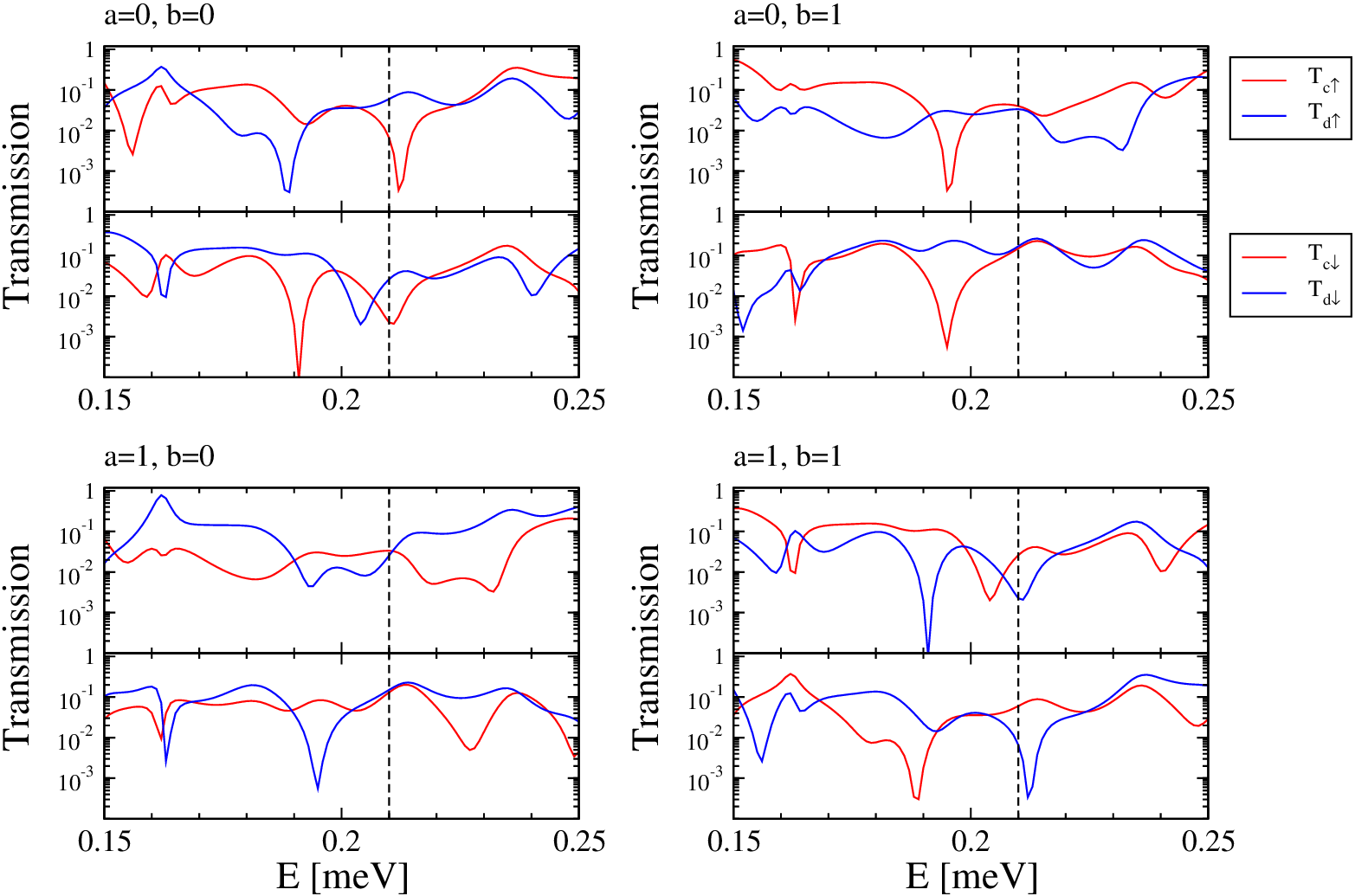} 
\caption{Transmission functions for a symmetric QPC showing the realization of an OR gate, at $E=0.21$ meV and $\alpha = 5.9$ meV nm.}
\label{or}
\end{figure}

\end{document}